\documentstyle{article}
\renewcommand{\vec}[1]{ \mbox{\boldmath $#1$} }  
\newcommand{\D}[2]{\frac{\partial #2}{\partial #1}}
\newcommand{\DD}[2]{\frac{\partial^2 #2}{\partial #1^2}}
\newcommand{\Dn}[3]{\frac{\partial^{#1} #3}{\partial #2^{#1}}}

\newcommand{\A}{{\cal A}}

\newcommand{\G}{{\cal G}}
\newcommand{\I}{{\cal I}}

\newcommand{\V}{{\cal V}}
\newcommand{\SetFigFont}[3]{}
\newcommand{\DITCH}[1]{}
\begin{document}
\title{\bf The construction of zonal models of dispersion in channels via
matched centre manifolds}
\author{S.D. Watt\thanks{Department of Mathematics, University of Melbourne,
Parkville, Victoria 3052, Australia. E-mail: {\tt sdw@mundoe.maths.mu.oz.au}}
\and A.J. Roberts\thanks {Department of Mathematics \& Computing,
University of Southern Queensland, Toowoomba, Queensland 4350, Australia.
E-mail: {\tt aroberts@usq.edu.au}}}
\date{November 28, 1994}
\maketitle

\begin{abstract}
Taylor's model of dispersion simply describes the long-term spread of
material along a pipe, channel or river. However, often we need multi-mode
models
to resolve finer details in space and time. Here we construct zonal models
of dispersion via the new principle of matching their long-term evolution
with that of the original problem. Using centre manifold techniques this is
done straightforwardly and systematically. Furthermore, this approach
provides correct initial and boundary conditions for the zonal models. We
expect the proposed principle of matched centre manifold evolution to be
useful in a wide range of modelling problems.
\end{abstract}

\section{Introduction}

This paper is an exploration and development of the principle of matched
centre manifolds in constructing low-dimensional models of dynamical
systems. By using the well studied example of shear dispersion in pipes and
channels, we demonstrate the utility of the new principle.

G.I.~Taylor\cite{Tay53} considered the dispersion of contaminant in a pipe.
He derived an advection-diffusion model for the longitudinal transport and
dispersion.
This model predicts a transport of contaminant with the average velocity
and with an effective diffusivity depending on the velocity profile and
cross-pipe diffusivity. Since then there have been a variety of approaches
to analysing dispersion, for example \cite{Smith,CO85,Gill}. There have also
been several studies \cite{MR90,WR94,MR94} using centre manifold theory to
derive such low-dimensional models of the dispersion; the basics are summarised
in Section~1.1. Centre manifold theory usefully provides these models with
initial and boundary conditions, and also caters for the presence of
spatial and temporal variations in the flow.

However, there are difficulties encountered in using these results to
predict contaminant dispersion. These problems include: restricted spatial
resolution, limited transient predictions, and difficulties in coding
high-order derivatives (especially in the boundary conditions). Many of
these problems are at least partially overcome by using an invariant
manifold approach, see \cite{WR94}, but such analysis is considerably more
difficult, especially for boundary conditions and for non-linear problems.

In this paper, we show how to overcome some of the difficulties. In
Section~2 we construct models of shear dispersion by requiring that a model
has the ``same'' centre manifold evolution, to some order, as that
for the original problem. This method of matching centre manifolds is a new
notion in the low-dimensional modelling of dynamical systems. (In some ways
it is similar to the idea of embedding a centre manifold, as discussed in
\cite{Rob92}.) In this application the constructed models may be called a
zonal as we identify a mode of slow advection with the near bank zone, and
a mode of fast advection~with the mid-stream zone.

Recently Chickwendu {\it et al} \cite{CO85} heuristically developed a
similar zonal model of dispersion in rivers and channels; the model
involved a mode to model the ``slow zone'' near the banks and bed, and a
mode to model the ``fast zone'' in the channel centre. However, a
difference is that here the parameters of the model are determined
systematically via centre manifold theory.

Furthermore, to construct a complete model we also need initial and
boundary conditions to supplement the evolution equation of the zonal
model. Appropriate initial conditions of the zonal model, given the initial
conditions of the original system, are found in Section~3 by matching the
initial conditions of both centre manifolds. Techniques described by
Roberts\cite{Rob9} give these initial conditions. Similarly, boundary
conditions for the centre manifold models, obtained using techniques
described in Roberts\cite{Rob92}, are matched in Section~4 to provide
correct inlet and outlet boundary conditions for the zonal model.

Thus this principle of matched centre manifolds systematically generates
models of arbitrary order complete with initial and boundary conditions.

\subsection{The centre manifold of channel dispersion}

As a prelude to this exploratory work, we here summarise the most basic
centre manifold model of shear dispersion in a channel.

Consider the flow of a contaminant in a channel of constant width, modelled
by the advection-diffusion equation \[ \D t{c} + \nabla\cdot(\vec q
c)=\nabla\cdot(\kappa \nabla c), \] where $c$ is the concentration of the
contaminant, $\vec q$ is the advection velocity and $\kappa$ is the
constant diffusivity. No flux of contaminant through the banks of the river
requires that $\D y{c}=0$ at $y=\pm b$, where $b$ is the half-channel
width.

We assume that the fluid is incompressible and the advection is along the
channel (the
$x$-direction) according to the velocity profile \[
\vec{q}=\vec{i}u(y)=\vec{i}\frac{3}{2}U\left(1-\frac{y^2}{b^2}\right), \]
where $U$ is the average velocity. As noted many times, see \cite{WR94} for
example, down-stream diffusion can be neglected without affecting more than
a few minor details. Doing this, $x$ and $y$ can be rescaled independently,
$y$ with respect to $b$ and $x$ with respect to $U b^2/\kappa$, so that in
effect $\kappa=1$ and $U=1$. Thus the non-dimensional equation to analyse
is \begin{equation} \D t{c} + \frac{3}{2} (1-y^2) \D x{c} = \DD y{c}.
\label{zm:origsys} \end{equation}

As the cross-stream diffusion operator $\DD y{}$ has one neutral mode and
all other modes decay, centre manifold techniques may be applied to analyse
the long-term behaviour of this system (as explained more fully by Mercer~\&
Roberts\cite{MR90}). The analysis is valid when the longitudinal
gradients, $\D x{}$, are small. Centre manifold theory \cite{Carr81}
assures us that the system (\ref{zm:origsys}) evolves exponentially quickly to
a low dimensional state which is dominated by the neutral mode. The system
then evolves slowly. To describe this low-dimensional, long-term evolution
Mercer \& Roberts\cite{MR90} assume that the system is
dependent only on this neutral mode, say
\begin{equation}
{c} = V(y,{C}) \quad\mbox{such that}\quad
\D t{{C}} = G({C})\,,\label{zm:cmans}
\end{equation}
where $C(x,t)$ is defined to be the cross-stream average of $c(x,y,t)$
and is therefore a measure of the ``amplitude'' of the neutral mode.

Mercer \& Roberts then developed asymptotic expansions
\begin{equation} V
\sim \sum_{n=0}^{\infty} v_{n}(y) \Dn n x{C}\mbox{~and~} G \sim
\sum_{n=1}^{\infty} g_{n} \Dn n x{C}\,,\label{zm:asymG}
\end{equation}
for these quantities where, for example,
\begin{eqnarray}
v_{0}(y) &=& 1\,,\\
g_{1} &=& -1\,,\label{zm:g1f}\\
v_{1}(y) &=& -\left(\frac{15 y^4 - 30 y^2 + 7}{120}\right)\,,\\
g_{2} &=& \frac{2}{105}\,,\label{zm:g2f}\\
v_{2}(y) &=&
\left(\frac{675 y^8 - 2940 y^6 + 3570 y^4 - 1020 y^2 -29}{201600}\right)\,,\\
g_{3} &=& \frac{4}{17325}\,.\label{zm:g3f}
\end{eqnarray}
These expansions were computed to high-order and shown to converge for
large scale structures, wave number $|k|<0.47$.

\section{Multi-mode models of channel dispersion}

We construct models of dispersion in channels (\ref{zm:origsys}) based on
the new proposed principle of matched centre manifolds. A multi-mode
low-order model is sought which has the same centre manifold evolution as
the original system---the agreement is to high-order. This principle is
rather like that employed in Pad\'e approximation (and related schemes
\cite[Chapt. 8]{BO}) where to improve the convergence of a Taylor series a
rational function is constructed which has the same Taylor series to a
specified order. Here we improve the resolution of a centre manifold model
by constructing a multi-mode model with the same long term evolution, to
some order in spatial derivatives.

Here we consider a two-component model. It eventuates that in effect one
component is slow and the other is fast as in the zonal models of
Chickwendu {\it et al} \cite{Chik86,CO85}. In this multi-mode model, we
posit conservative exchange between the modes, and advection and diffusion
within each mode, but no intermodal advection or diffusion. This is shown
schematically in Figure~\ref{zm:flow}, where the two modes are interpreted
as two zones of different ``widths'' or ``capacities''. A model system with
these properties is
\begin{equation}
\D t{\vec u} = A \vec{u} - B \D x{\vec{u}} + D \DD x{\vec{u}}\,, \label{zm:icz}
\end{equation}
where ${\vec u}=(u_1,u_2)$,
and where
\[
A= \left[\begin{array}{cc}
-a&a\\
b&-b\end{array}\right],\qquad
B = \left[\begin{array}{cc}
s_{1}&0\\
0&s_{2}\end{array}\right],\qquad
D = \left[\begin{array}{cc}
d_{1}&0\\
0&d_{2}\end{array}\right].
\]
Observe that $bu_1+au_2$ is conserved; if $u_1$ and $u_2$ are considered
to be ``concentrations'' in the zones, then the ``width'' of the zones is
in the ratio of $b:a$ (see Figure~\ref{zm:flow}).

To determine the six constants of this model we match the long-term
evolution of (\ref{zm:origsys}), as occurs on the centre manifold
(\ref{zm:asymG}), with the long-term evolution of this model, as expressed
on its centre manifold.

\subsection{The centre manifold of the model}

As before, to construct the centre manifold we take derivatives with
respect to $x$ be a small parameter. Dominantly, then
\[
\D t{\vec{u}}=A\vec{u}\,,
\]
and so the ``concentrations'' equilibrate between the zones with transients
approximately like $e^{-(a+b)t}$. The centre manifold is thus described in
terms of the cross-zone weighted-average $C = ({b u_{1} + a u_{2}})/({a+b})$.
As usual \cite{Rob8}, the centre manifold is constructed by assuming
\[
\vec{u}=\vec{\V}(C) \quad\mbox{such that}\quad \D t C=\G(C) \,,
\]
and then seeking asymptotic expansions for $\V$ and $\G$ of the form
\[
\vec{\V}\sim\sum_{n=0}^{\infty} \vec{v}_{n} \Dn n x{C}
\quad\mbox{and}\quad
 \G\sim\sum_{n=1}^{\infty} g_{n} \Dn n x{C}\,. \]
Substituting these into the model equation (\ref{zm:icz}), and collecting
like longitudinal derivatives $\Dn n x C$, gives a hierarchy of equations
\begin{figure}
\begin{center}
\input{scheme2.latex}
\end{center}
\caption{A schematic representation of the zonal model showing the three
mechanisms of the channel---exchange, advection and diffusion.}
\label{zm:flow}
\end{figure}
\begin{equation}
A {\vec v}_{n} = \sum_{m=1}^{n} {\vec v}_{n-m} g_{m} + B {\vec v}_{n-1} - D
{\vec v}_{n-2}\,.
\label{zm:hier}
\end{equation}

These together with amplitude conditions are easily solved to high-order
via the same form of {\sc reduce} program as used for the original system.
For example, to second order the evolution on the centre manifold is \[ \D
t{C} \sim -\overline{u} \D x{C} + \overline{d}\DD x{C}, \]
where
\begin{equation}
\overline{u}=\frac{b s_{1} + a s_{2}}{a+b}\,, \label{zm:g1z}
\end{equation}
is the appropriately weighted mean advection velocity, and the effective
diffusivity is
\begin{equation}
\overline{d}=\frac{ab(s_{1}-s_{2})^2}{(a+b)^3}+\frac{b d_{1} + a
d_{2}}{a+b}\,. \label{zm:g2z}
\end{equation}
This effective diffusivity is the superposition of the weighted mean
diffusivity, $({b d_{1} + a d_{2}})/({a+b})$, and the shear dispersion
term, ${ab(s_{1}-s_{2})^2}/{(a+b)^3}$, which is proportional to the
square of the velocity difference in the two zones. Using {\sc reduce} we
easily compute higher-order terms in the expansions.

\subsection{Matching for the advection model}

The objective is to find a zonal model (\ref{zm:icz}) whose evolution is
``close'' to the original system (\ref{zm:origsys}) of shear dispersion. Thus
we must find good parameters to make the connection. A straightforward way,
given that the evolution on both centre manifolds is known, is to assert
that the long-term evolution on each manifold is the same to some order. As
there are six degrees of freedom in the zonal model, those being the as yet
undetermined parameters $a$, $b$, $s_{1}$, $s_{2}$, $d_{1}$ and $d_{2}$, we
determine an agreement up to sixth order in $\D x{}$.

First, for comparison, we match both models for the case where there is no
diffusion in the zonal model ($d_1=d_2=0$). This reduces the number of
parameters by two and so we only seek agreement to fourth order.

Equating the coefficients of the first four derivatives of each evolution
equation, namely (\ref{zm:g1z}) with (\ref{zm:g1f}), (\ref{zm:g2z}) with
(\ref{zm:g2f}) and so on, gives four non-linear equations in four unknowns.
These equations have the solution \begin{eqnarray*}
a&=&\frac{4719}{812}-\frac{4719}{481516}\sqrt{7709}\approx 4.9511\,,\\
b&=&\frac{4719}{812}+\frac{4719}{481516}\sqrt{7709}\approx 6.6721\,,\\
s_{1}&=&\frac{1887}{2030}+\frac{11}{2030}\sqrt{7709}\approx 1.4053\,,\\
s_{2}&=&\frac{1887}{2030}-\frac{11}{2030}\sqrt{7709}\approx 0.4539\,.
\end{eqnarray*}
As the parameters $a$ and $b$ are in essence the capacity
of each zone, these result suggest that the fast zone should be thought of
as nearly one and a half times as wide as the slow zone. Physically, we may
imagine that the fast-zone occupies the middle three-fifths of the channel,
whereas the slow-zone corresponds to the two outer fifths. However, this
identification is refined in Section~3.4 when initial conditions are found.

Also, the second eigenvalue of the interaction matrix $A$, approximately
the decay rate onto the centre manifold, is $-(a+b)$. Here this is
$-\frac{4719}{406}\approx -11.62$, which is comparable (perhaps
fortuitously, but discussed later) to $-\pi^2\approx -9.8696$, the decay
rate of the first neglected symmetric mode of the original
system~(\ref{zm:origsys}), a difference of about $18\%$. Thus, not only is
the long-term evolution nearly identical, but also the rate of approach to
the shared long-term evolution is similar in both the model and the
original.

\subsection{Matching with diffusion}

Including the diffusion term in the zonal model makes the matching more
difficult. Thus the equations are not solved analytically, but numerically.
The obtained solution has $a=4.5669$ and $b=5.6569$ and $s_1$, $s_2$,
$d_1$ and $d_2$ listed in Table~\ref{table}. For comparison
the coefficients found by Chickwendu\cite{Chik86} and the previous
diffusionless model are also listed.  The
coefficients of the three models are very similar.
\begin{table}\begin{center}
\begin{tabular}{|c|c|c|c|} \hline
&Without diffusion&With diffusion&Chickwendu\\
\hline $\eta_1$&0.5740&0.5533&0.5774\\
$\eta_2$&0.4260&0.4467&0.4226\\
$s_1$&1.4053&1.3829&1.3333\\
$s_2$&0.4539&0.5257&0.5447\\
$d_1$&-&0.5043$\times 10^{-3}$&$0.7055\times 10^{-3}$\\
$d_2$&-&2.2426$\times 10^{-3}$&$1.4903\times 10^{-3}$\\\hline
\end{tabular}
\end{center}
\caption{A comparison of the parameters for each of the three zonal models,
where $\eta_1={b}/({a+b})$ and $\eta_2={a}/({a+b})$ as used by Chikwendu.}
\label{table}
\end{table}

Note the presence of a pleasing physical feature in this
zonal model. The low value of effective diffusivity in the fast zone,
$d_1$, neatly corresponds to the limited shear of the fast flow in the
centre of the channel; whereas the comparatively high value in the slow
zone, $d_2$, matches the high shear found near the channel banks.

Lastly, the decay rate onto the centre manifold, $-(a+b)=-10.2238$, is again
remarkably close to $-\pi^2$, a difference of about $3.6\%$.  This
closeness may be explained by noting that a centre manifold
analysis is similar to that of a perturbed eigenproblem where here the
derivative, $\D x{}$, is the perturbing parameter. Typically, in a perturbed
eigenproblem the different eigenvalues are analytic continuations of each
other and are identifiable as different branches, or Riemann sheets, of the
one analytic function \cite[Section~7.5]{BO}. Thus the expansion for any
one eigenvalue, here the neutral mode corresponding to the centre manifold,
is affected by the other eigenvalues through their continuation in the
complex plane. Hence it is plausible to expect that the exponentially
decaying transients of the model, here dominantly $\exp[-(a+b)t]$, do
correspond to physical dynamics in the original problem. However, due to the
symmetry of the original problem about the channel centreline, the
symmetric and antisymmetric Riemann sheets are completely decoupled. Hence
the zonal model can only be affected by the symmetric channel modes.

\subsection{A comparison of our two zonal models}

There are relatively minor differences between the widths and advection
velocities for the models with and without diffusion: with diffusion the
fast zone is a little ``thicker'' and slower while the slow zone is a
little ``smaller'' and faster. For a further comparison between our two
zonal models, we investigate how well the
evolution on the centre manifold agrees between the two zonal models and
the original system. Consider the asymptotic expansions for the evolution on
the centre manifolds, of the form
\[
G\sim\sum_{n=1}^{\infty} g_n \Dn n x{C}\,. \]
As discussed in Mercer \& Roberts \cite[Appendix]{MR90}, the validity of
these expansions is related to the radius of convergence of the series. By
using the Roberts' generalisation of the Domb-Sykes formula
\[
B_k^2=\frac{g_{k+1}g_{k-1}-g_{k}^2}{g_{k}g_{k-2}-g_{k-1}^2}\,, \]
the radius
of convergence is given as
\[
\frac{1}{r_c}=\lim_{k\rightarrow\infty} B_k\,. \]
Plotting $B_k$ versus $1/k$, we extrapolate to find the radius of
convergence, $r_c$, of each expansion. For the zonal model without
diffusion, the radius of convergence is about $12.15$. For the model with
diffusion and the original system, both have a radius of convergence of about
$11.8$. (See Figure~\ref{zm:dsplot}; note that the original system and the
diffusive zonal model are almost indistinguishable.)
\begin{figure}
\caption{Generalised Domb-Sykes plot of the series of the $B_k$'s, for the
original system and both zonal models, showing the close match of the evolution
on the centre manifolds at high order. The original system is the dotted
circle, the zonal model with diffusion is the empty squares, and the zonal
model without diffusion is the filled squares.}
\label{zm:dsplot}
\end{figure}

This shows that by just matching evolutions at low order, important
properties of the evolution on the centre manifold, such as convergence,
are also closely matched at higher order: the model without diffusion is a
good approximation; the model with diffusion an even closer approximation.
However, this does not imply that the models will give identical behaviour
inside their common radii of convergence.  But it does imply that
higher-order corrections to the zonal models are likely to be small
because the high-order behaviour is already closely matched to that of
the centre manifold evolution of the original system.

\section{Initial conditions}

By centre manifold theory \cite{Carr81}, for every trajectory starting near
the centre manifold there is guaranteed to bea specific solution on the
centre manifold which is approached exponentially quickly. Thus, for any
initial distribution of contaminant, $c^{0}$, of the original system, there is
an initial condition, $C^0$, for starting on the manifold so that the
centre manifold solution approaches the exact solution exponentially
quickly. Traditionally it has been assumed that $C^0$ is just the initial
cross-sectional average of the concentration. This is roughly correct, but
is initially in error, and the errors persist for all time. Dynamically
based arguments to derive the correct initial condition, given an original
exact initial condition, were described by Roberts\cite{Rob9} for a general
nonlinear system. For a linear system, such as the dynamics of shear
dispersion, the derivation may be simplified as described by Watt \&
Roberts \cite[Section~3]{WR94}. It is the later formulation adopted here.

However, here we need to find an appropriate projection from the initial
condition of the original system to the zonal model, as indicated by the dotted
line labelled $\vec \zeta$ in Figure~\ref{zm:ic}. This is found using the
projection from the original system to the centre manifold, $z$ on the figure,
and that from the zonal model to the centre manifold, $\vec Z$ on the
figure. Then $\vec \zeta$ is determined by requiring that the composition
of the projection from the original system to the zonal and thence onto the
centre manifold, is the same as that directly from the original system to the
centre manifold; that is we find $\vec \zeta$ so that
$\vec{Z}\circ\vec\zeta$ is the same, to some order, as $z$. To obtain the
correct initial condition for the centre manifolds (solid arrows in
Figure~\ref{zm:ic}) we use the arguments and formulae developed by us
\cite[Section~3.1]{WR94}, and summarised below.
\begin{figure}
\caption{Diagram showing how the initial conditions of the zonal model are
found given the initial condition of the original system, done by matching the
initial condition on both centre manifolds.}
\label{zm:ic}
\end{figure}

\subsection{Summary of the general linear analysis}

Consider a general linear system
\begin{equation}
\dot{\vec u} = \A \vec u\,,
\end{equation}
where $\A$ is some particular linear operator (implicitly including boundary
conditions if a differential operator), and the evolution on a
low-dimensional invariant subspace $\vec u = \V \vec c$, where $\V$ is a
linear operator spanning the subspace. To be invariant under the
evolution, this subspace must be spanned by a set of eigenvectors of $\A$.
The evolution on the subspace may then be described by some low-dimensional
evolution equation $\dot{\vec c} = \G \vec c$

As previously argued \cite[Section~3]{WR94}, there is a projection
operator which will take any solution of the original system down onto a
solution on the manifold, namely, that solution on the manifold which is
approached exponentially. This operator $(\vec{Z},\ldots)$ may be expressed
as
\begin{equation}
\left( \vec{Z} , \vec{u}(t) \right) = \vec{c}(t)\,, \label{ic0}
\end{equation}
for some inner product. For example, in the original system we
use the inner product
\begin{equation}
\left( u , v \right) = \frac{1}{2}\int_{-1}^{1} u v\, dy=\overline{u v}\,,
\end{equation}
and in the zonal model, use
\begin{equation}
\left( \vec{u},\vec{v} \right ) = \vec{u}^{T} \vec{v}\,. \end{equation}
{}From \cite{WR94}, the projection $\vec Z$ is the solution of
\begin{equation}
\A^\dagger \vec{Z}= \vec{Z}\G^T\,,
\label{ic1}
\end{equation}
and the orthogonality equation
\begin{equation}
\left( \vec{Z},\V \right) = \I\,.
\label{ic2}
\end{equation}

\subsection{Initial condition from the original system}

For the centre manifold of the original system, (\ref{zm:origsys}), we identify
\[
z\sim \sum_{n=0}^{\infty} z_{n}(y) \Dn n x{}
\,,\quad
\G \sim \sum_{n=1}^{\infty} g_{n} \Dn n x{}
\,,\quad
\A = {\cal L} + u(y) \D x{}\,,
\]
where ${\cal L}=\DD y{}$ with boundary conditions of no flux across the
channel boundaries.

Substituting these into equation~(\ref{ic1}) and collecting terms of the
same order together yields
\[
{\cal L} z_{n} = \sum_{m=1}^{n} z_{n-m} g_{m} + u(y) z_{n-1}\,. \]
Since $\cal L$ is self-adjoint, this is in exactly the same form as the
equation (2.12) in \cite{MR90}, which was solved for the centre manifold of
the original system. There are also the subsidiary conditions
\[
\overline{z_0} = 1\quad\mbox{and}\quad
\sum_{m=0}^{n}\overline{z_{n-m}v_{m}}=0\,, \]
as a consequence of the orthogonality constraint (\ref{ic2}).

To the first few orders, the initial condition is
\begin{equation}
C^{0}\sim\overline{z_{0} c^{0}}+\overline{z_{1} \D x{c^{0}}}
+\overline{z_{2} \DD x{c^{0}}}+\overline{z_3 \Dn 3 x{c^{0}}}\,,
\label{zm:cmic}
\end{equation}
where
\begin{eqnarray*}
z_{0}(y)& =& 1\,,\\
z_{1}(y)&=&-\frac{15 y^4 - 30 y^2 + 7}{120}\,,\\
z_{2}(y)&=&\frac{675 y^8-2940 y^6+3570 y^4-1020 y^2 - 413}{201600}\,,\\
z_{3}(y)&=&-\frac{675y^{12}-4642y^{10}+10725y^8-8316y^6}{17740800}\\
&&+\frac{10705695y^4-23060310y^2+4076777}{24216192000}\,.
\end{eqnarray*}
as recorded by Mercer \& Roberts \cite{MR90}. Higher orders were also
computed in order to perform the matching.

\subsection{Initial condition from the zonal model}

For the zonal model we identify
\[
\vec{Z} \sim \sum_{n=0}^{\infty} \vec{Z}_{n} \Dn n x{}\,,
\quad
 \G \sim \sum_{n=1}^{\infty} g_{n} \Dn n x{}\,,
 \quad
  \A = A - B \D x{} + D \DD x{}\,.
\]
Substituting these into equation~(\ref{ic1}) and collecting terms of the
same order together, yields
\[
A^{T} \vec{Z}_{n} = \sum_{m=1}^{n} \vec{Z}_{n-m} g_{m} + B^{T}
\vec{Z}_{n-1} - D^{T} \vec{Z}_{n-2}\,.
\]
As can be seen, this is in the same form as equation~(\ref{zm:hier}). The
differences are that the matrices are transposed and that now an
orthogonality constraint,
\[ \vec{Z}^{T}_{0}\vec{v}_{0} = 1 \quad\mbox{and}\quad
\sum_{m=0}^{n} \vec{Z}^{T}_{n-m}\vec{v}_{m} = 0\,, \]
needs to be satisfied.

Solving this hierarchy of equations to the first few orders, we deduce that
the initial condition for the centre manifold of the zonal model is
approximately
\begin{equation}
C^{0} \sim \vec{Z}^{T}_{0}\vec{u}^{0} + \vec{Z}^{T}_{1} \D x{\vec{u}^{0}}
+ \vec{Z}^{T}_{2} \DD x{\vec{u}^{0}}+\vec{Z}^{T} \Dn 3 x{\vec{u}^0}\,,
\label{zm:zmic3}
\end{equation}
where
\begin{eqnarray*}
\vec{Z}_{0} &=& \frac{1}{a+b}\left[
\begin{array}{c}
b\\a
\end{array} \right],\\
\vec{Z}_{1} &=& \frac{ab(s_1-s_2)}{(a+b)^3}\left[ \begin{array}{c}
-1\\1\end{array} \right],\\
\vec{Z}_{2} &=& \frac{ab(s_1-s_2)^2}{(a+b)^5}\left[ \begin{array}{c}
a-2b
\\b-2a
\end{array} \right]+
\frac{ab(d_1-d_2)}{(a+b)^3}\left[\begin{array}{c}1\\-1\end{array}\right],\\
 \vec{Z}_3&=& \frac{ab(s_1-s_2)^3}{(a+b)^7}
 \left[\begin{array}{c}-a^2+6ab-3b^2\\3a^2-6ab+b^2\end{array}\right]
 + \frac{ab(s_1-s_2)(d_1-d_2)}{(a+b)^5}
\left[\begin{array}{c}a-2b\\2a-b\end{array}\right].
\end{eqnarray*}
Higher orders were also calculated to be used for the matching.

\subsection{Matching without diffusion}

We now find the $\vec{u}^{0}$ for which the zonal model matches a given
$c^{0}$ by equating the expressions~(\ref{zm:cmic}) and~(\ref{zm:zmic3})
for the two initial conditions found on the centre manifold.

Suppose $\vec{u}^{0}$ is given by a projection of the form
\begin{equation}
\vec{u}^{0} =
\overline{\vec{\zeta}_{0}(y) c^{0}}
+ \overline{\vec{\zeta}_{1}(y) \D x{c^{0}}}\,,
\label{zm:u0wd}
\end{equation}
where the as yet unknown $2\times 1$ matrices $\vec{\zeta}_{0}$ and
$\vec{\zeta}_1$ are to be determined by matching. Now the initial condition
on the centre manifold direct from the original system is~(\ref{zm:cmic}),
whereas that for the centre manifold of the zonal model after the as yet
unknown projection~(\ref{zm:u0wd}) from the original system onto the model is
\begin{eqnarray*}
C^{0}= \overline{\vec{Z}^{T}_{0}\vec{\zeta}_{0} c^{0}} +
\overline{\vec{Z}^{T}_{0}\vec{\zeta}_{1} \D x{c^{0}}}&& \nonumber\\ +
\overline{\vec{Z}^{T}_{1}\vec{\zeta}_{0} \D x{c^{0}}} &+&
\overline{\vec{Z}^{T}_{1}\vec{\zeta}_{1} \DD x{c^{0}}} \\ &+&
\overline{\vec{Z}^{T}_{2}\vec{\zeta}_{0}\DD x{c^{0}}} +
\overline{\vec{Z}^{T}_{2} \vec{\zeta}_{1} \Dn 3 x{c^{0}}} + \cdots \,.
\nonumber \end{eqnarray*}
These two expressions for $C^0$ must be equal for all initial distributions
$c^0$, so we equate coefficients of $c^0$ and its derivatives. Equating the
four integrands up to $3^{rd}$ order, we get four scalar equations in the
unknown $\vec{\zeta}_0$ and $\vec{\zeta}_1$:
\begin{eqnarray*}
z_{0}(y)&=&\vec{Z}^{T}_{0}\vec{\zeta}_{0}\,,\\
z_{1}(y)&=&\vec{Z}^{T}_{0}\vec{\zeta}_{1} + \vec{Z}^{T}_{1}
\vec{\zeta}_{0}\,,\\
z_{2}(y)&=&\vec{Z}^{T}_{1} \vec{\zeta}_{1} + \vec{Z}^{T}_{2}
\vec{\zeta}_{0}\,,\\
z_{3}(y)&=&\vec{Z}^{T}_{2} \vec{\zeta}_{1} + \vec{Z}^{T}_{3} \vec{\zeta}_{0}\,.
\end{eqnarray*}
Solving these linear equations gives
\begin{eqnarray*}
\vec{\zeta_{0}} &=&
\left[\begin{array}{c}1.2580\\0.6524\end{array}\right]z_{0}
+ \left[\begin{array}{c}21.28\\-28.68\end{array}\right]z_{1}
+ \left[\begin{array}{c}-157.40\\212.0\end{array}\right]z_{2}
+ \left[\begin{array}{c}6493\\-17498\end{array}\right]z_{3}\,,\\
 \vec{\zeta_{1}} &=&
\left[\begin{array}{c}-0.03588\\0.06332\end{array}\right]z_{0}
+ \left[\begin{array}{c}1.7420\\2.348\end{array}\right]z_{1}
+ \left[\begin{array}{c}-26.76\\18.708\end{array}\right]z_{2}
+ \left[\begin{array}{c}452.8\\822.4\end{array}\right]z_{3}\,.
\end{eqnarray*}
These functions are shown graphically in Figure~\ref{zm:icwod}. Observe
from Figure~\ref{zm:icwod}(a) that any contaminant released near the
channel centre is assigned to the fast zone mode, whereas any released near
the banks is assigned to the slow zone mode. However, there is no sharp
boundary between the physical zones, the transition is smooth. Also observe
that the corrections $\vec{\zeta}_1$, shown in Figure~\ref{zm:icwod}(b),
are about 1\% of $\vec{\zeta}_0$, hence only steep gradients in the
initial concentration alter the
leading order in the initial condition~(\ref{zm:u0wd}).
\begin{figure}
\caption{graphs of the initial condition functions as a function of $y$ for
the zonal model without diffusion: (a) ${ \zeta}_0$; (b) ${ \zeta}_1$. The
solid line (------) is for the fast zone, ${ \zeta}_{n1}$, and the dotted
line ($\cdots\cdots\cdots$) is for the slow zone, ${ \zeta}_{n2}$}
\label{zm:icwod}
\end{figure}

\subsection{Matching with diffusion}

Following the same method as in the previous subsection, we find
$\vec{u}^{0}$ given $c^{0}$, except now we suppose $\vec{u}^{0}$ is of the form
\[
\vec{u}^{0} = \overline{\vec{\zeta}_{0}(y) c^{0}}
+ \overline{\vec{\zeta}_{1}(y) \D x{c^{0}}}
+ \overline{\vec{\zeta}_{2}(y) \DD x{c^{0}}}\,.
\]
Equating the integrands up to $5^{th}$ order, we get six equations in the
unknown $\vec{\zeta}_n$ functions:
\begin{eqnarray*}
z_{0}(y)&=&\vec{Z}^{T}_{0} \vec{\zeta}_{0}\,,\\
z_{1}(y)&=&\vec{Z}^{T}_{0} \vec{\zeta}_{1} + \vec{Z}^{T}_{1}
\vec{\zeta}_{0}\,,\\
z_{2}(y)&=&\vec{Z}^{T}_{0} \vec{\zeta}_{2} + \vec{Z}^{T}_{1} \vec{\zeta}_{1}
+ \vec{Z}^{T}_{2} \vec{\zeta}_{0}\,,\\
z_{3}(y)&=&\vec{Z}^{T}_{1} \vec{\zeta}_{2} + \vec{Z}^{T}_{2} \vec{\zeta}_{1}
+ \vec{Z}^{T}_{3} \vec{\zeta}_{0}\,,\\
z_{4}(y)&=&\vec{Z}^{T}_{2} \vec{\zeta}_{2} + \vec{Z}^{T}_{3} \vec{\zeta}_{1}
+ \vec{Z}^{T}_{4} \vec{\zeta}_{0}\,,\\
z_{5}(y)&=&\vec{Z}^{T}_{3} \vec{\zeta}_{2} + \vec{Z}^{T}_{4} \vec{\zeta}_{1}
+ \vec{Z}^{T}_{5} \vec{\zeta}_{0}\,.
\end{eqnarray*}
Solving these linear equations, we find
\begin{eqnarray*}
\vec{\zeta}_0(y)&=&
\left[\begin{array}{c}1.4928\\0.4144\end{array}\right]z_{0}+
\left[\begin{array}{c}21.56\\-26.70\end{array}\right]z_{1}+
\left[\begin{array}{c}-0.3796\\0.4702\end{array}\right]z_{2}\\&&+
\left[\begin{array}{c}-25636\\31756\end{array}\right]z_{3}+
\left[\begin{array}{c}156000\\-96616\end{array}\right]z_{4}+
\left[\begin{array}{c}-7121926\\8821738\end{array}\right]z_{5}\,,\\
\vec{\zeta}_1(y)&=&
\left[\begin{array}{c}-0.06074\\0.11436\end{array}\right]z_{0}+
\left[\begin{array}{c}1.7638\\2.292\end{array}\right]z_{1}+
\left[\begin{array}{c}21.50\\-26.66\end{array}\right]z_{2}\\&&+
\left[\begin{array}{c}-763.6\\-1716.8\end{array}\right]z_{3}+
\left[\begin{array}{c}16194\\3856\end{array}\right]z_{4}+
\left[\begin{array}{c}-174428\\-523658\end{array}\right]z_{5}\,,\\
\vec{\zeta}_2(y)&=&
\left[\begin{array}{c}-0.0010058\\-0.002914\end{array}\right]z_{0}+
\left[\begin{array}{c}0.00009333\\-0.00010574\end{array}\right]z_{1}+
\left[\begin{array}{c}1.7612\\2.290\end{array}\right]z_{2}\\&&+
\left[\begin{array}{c}-61.08\\90.70\end{array}\right]z_{3}+
\left[\begin{array}{c}649.6\\303.2\end{array}\right]z_{4}+
\left[\begin{array}{c}-9646\\20038\end{array}\right]z_{5}\,.\\
\end{eqnarray*}
\begin{figure}
\caption{graphs of the initial condition functions as a function of $y$ for
the zonal model with diffusion: (a) ${\zeta}_0$; (b) ${\zeta}_1$; (c)
${\zeta}_2$. The solid line (---------) is for the fast zone,
${\zeta}_{n1}$, and the dotted line ($\cdots\cdots\cdots$) is for the slow
zone, ${\zeta}_{n2}$.}
\label{zm:icwd}
\end{figure}
These functions are shown graphically in Figure~\ref{zm:icwd}. As in
Figure~\ref{zm:icwod}(a), most of the contribution to the initial condition
of the fast zone comes from the middle of the channel, the location of the
fast zone, and most of the initial condition of the slow zone comes from
near the banks, the location of the slow zone.

\subsection{Comparative results}

Using these initial conditions for both zonal models, we compare the
original
system with the approximate models via some numerical simulations. As well
as comparing the two zonal models with the original system~(\ref{zm:origsys}),
we also include predictions for the centre manifold model~(\ref{zm:cmans})
with initial condition~(\ref{zm:cmic}).

The numerical simulations employed simple finite difference schemes for
each model and the original system. The channel is long enough so that neither
the inlet nor the outlet had any influence on the contaminant field.

The initial contaminant field chosen for the comparison was a mid-channel
release of the form
\[
c^0(x,y)=\exp\left[{-(2x)^{12}-(4y)^{12}}\right],
\]
which approximates a box of length $1$ and width $0.5$ at the centre of the
channel. The property chosen to base the comparison on was the average
concentration across the channel, as a function of downstream position.
This is shown in Figure~\ref{zm:comp} at time $t=0.1$.
\begin{figure}
\caption{Comparison of the mean concentration of each model at time
$t=0.1$ where: the centre manifold model~(\protect{\ref{zm:cmans}}) is the
solid line (---------); the zonal model without diffusion is the short-dashed
line (- - - - - -); the zonal model with diffusion is the dashed line
(-- -- -- --); and the original system~(\protect{\ref{zm:origsys}}) is shown by
the discs ({\protect\tiny $\bullet$}). Each model simulation started from the
initial conditions determined in this section.}
\label{zm:comp}
\end{figure}

{}From this figure, it can be seen that both zonal models are very good
approximations for this small time, indeed they are both nearly
indistinguishable from the exact solution, with the zonal model with
diffusion better than the model without diffusion (see inset).
Importantly, the matching process used to guarantee a long-term agreement
between model and original, here also produces excellent short-term
predictions.

Observe that all three models predict a concentration which is negative in
a very
small region at the ``tail'' of the profile, near $x\approx -0.5$. This is
due to the corrections of the initial conditions and the fact the
contaminant is conserved. As shown in \cite{WR94} such negative
concentrations are a necessary condition for long-term agreement between
model and original system.

We also ran the model to obtain solutions for time $t=1$ to show a little
of the long-term agreement between the dynamics. In
Figure~\ref{diff} the errors in the zonal models are shown to be typically
less than $10^{-4}$; with the higher-order diffusion model being the
better.   The centre manifold model has errors which are two orders
of magnitude larger; still small because correct initial conditions
ensure a long-term agreement, but not as good as the zonal models.
\begin{figure}
\caption{the log (to base 10) of the difference between a model and
the original system at time $t=1$: the centre manifold model of \S\S1.1 is
the solid line (--------); the zonal model without diffusion of \S\S2.2 is
the small-dashed line (- - - -); and the zonal model with diffusion of \S\S2.3
is the dashed line (-- -- --). Each model simulation started from the
initial conditions determined in this section.}
\label{diff}
\end{figure}

\section{Boundary conditions}

The zonal models~(\ref{zm:icz}) are partial differential equations in space
and time. Hence spatial boundary conditions need to be specified before the
equations are solved. We find boundary conditions via adaptations of the
method developed by Roberts~\cite{Rob92a} for centre manifold models.

\subsection{Inlet boundary conditions}

First the inlet boundary condition of the zonal model is found as a
function of the inlet boundary condition of the original system. As explained
in \cite{Rob92a}, finding the appropriate boundary condition on the centre
manifold is similar to that of finding the initial condition on the centre
manifold, except that here the governing equations are taken to describe
the evolution in space given slow time variations. The ``initial''
condition of the spatial evolution is equivalent to the inlet boundary
condition of the time evolution.

The equations for both the original system and the zonal model (without
diffusion) are rewritten respectively as
\begin{eqnarray*}
{\cal L}c&=&u(y)\D x{c}+\D t{c}\,,\\
A\vec{u}&=&B\D x{\vec{u}}+\D t{\vec{u}}\,,
\end{eqnarray*}
which are (\ref{zm:origsys}) and (\ref{zm:icz}) (without the diffusion
term), except that here the ``advection'' coefficients, $u(y)$ and $B$
respectively, multiply what we now consider as the ``time-like''
derivative.

We perform the same analysis as in Sections~1 and~2 to get
the approximation to the centre manifolds of the spatial evolution
\begin{eqnarray*}
c&\sim&w_0C+w_1\D t{C}+w_2\DD t{C}+\cdots\,,\\
\vec{u}&\sim&\vec{w}_0C+\vec{w}_1\D t{C}+\vec{w}_2\DD t{C}+\cdots\,,
\end{eqnarray*}
where
\begin{eqnarray*}
w_0(y)&=&1\,,\\
w_1(y)&=&\frac{15y^4-30y^2+7}{120}\,,\\
w_2(y)&=&\frac{675y^8-2940y^6+3090y^4-60y^2-253}{201600}\,,
\end{eqnarray*}
and
\begin{eqnarray*}
\vec{w}_0&=&\left[\begin{array}{c}1\\1\end{array}\right],\\
\vec{w}_1&=&\frac{s_1-s_2}{(a+b)(as_2+bs_1)}
\left[\begin{array}{c}a\\-b\end{array}\right],\\
\vec{w}_2&=&\frac{(s_1-s_2)^2(a^2s_2-b^2s_1)}{(a+b)^2(as_2+bs_1)^3}
\left[\begin{array}{c}a\\-b\end{array}\right],
\end{eqnarray*}
where the evolution on the centre manifold of the original system is
\begin{equation}
\D x{C}\sim -\D t{C} +\frac{2}{105}\DD t{C}+\cdots\,,
\label{zm:xevol}
\end{equation}
and the evolution on the centre manifold of the zonal model is
\begin{equation}
\D x{C}\sim-\overline{u}' \D t{C}+\overline{d}'\DD t{C}+\cdots\,,
\label{zm:xevolzm}
\end{equation}
where
\[
\overline{u}'=\frac{a+b}{as_2+bs_1}
\quad\mbox{and}\quad
 \overline{d}'=\frac{ab(s_1-s_2)^2}{(as_2+bs_1)^3}\,.
\]
By equating the coefficients of the first four derivatives in the evolution
equation~(\ref{zm:xevol}) with the first four coefficients
in~(\ref{zm:xevolzm}), we find that the parameters of the zonal model $a$,
$b$, $s_1$ and $s_2$ are exactly the same as those determined in
Section~2.2. As the temporal and spatial evolution equations are closely
related, the reversion of each other, it would be expected that the
parameters will be the same. Thus, this is a useful confirmation that the
derived evolution equations are correct, but does not give any new
information.

By following the method outlined in Section~3, we find the inlet boundary
condition of the centre manifold corresponding to the inlet boundary
condition of the original system and the zonal model. The inner product used is
a weighted average with respect to the velocity, $u(y)$ and $B$
respectively, that is
\[
\langle\alpha,\beta\rangle=\frac{1}{2}\int_{-1}^{1} u(y)\alpha(y)\beta(y)\, dy
=\overline{u\alpha\beta}\,,
\]
is the inner product for the original system, and
\[
\langle\vec{\alpha},\vec{\beta}\rangle=\vec{\alpha}^TB\vec{\beta}\,,
\]
is the inner product for the zonal model.
The inlet condition for the centre manifold of the original system
is found to be
\begin{equation}
C(0,t)\sim\overline{up_0c(0,t)}+\overline{up_1\D t{c}(0,t)}
+\overline{up_2\DD t{c}(0,t)}+\cdots\,,
\label{zm:cmbc}
\end{equation}
where
\begin{eqnarray*}
p_0(y)&=&1\,,\\
p_1(y)&=&\frac{105y^4-210y^2+17}{840}\,,\\
p_2(y)&=&\frac{51975y^8-226380y^6+164010y^4+143220y^2-39001}{15523200}\,.
\end{eqnarray*}
whereas the inlet condition of the centre manifold corresponding to the
zonal model is
\begin{equation}
C(0,t)\sim\vec{P}_0^TB\vec{u}(0,t)+\vec{P}_1^TB\D t{\vec{u}}(0,t)
+\vec{P}_2^TB\DD t{\vec{u}}(0,t)+\cdots\,,
\label{zm:u0bc}
\end{equation}
where
\begin{eqnarray*}
\vec{P}_0&=&\frac{1}{as_2+bs_1}\left[\begin{array}{c}b\\a\end{array}\right],\\
\vec{P}_1&=&\frac{ab(s_1-s_2)}{(a+b)(as_2+bs_1)^3}
\left[\begin{array}{c}as_2-2as_1-bs_1\\as_2+2bs_2-bs_1\end{array}\right],\\
\vec{P}_2&=&\frac{3ab(s_1-s_2)^2}{(a+b)(as_2+bs_1)^5}
\left[\begin{array}{c}bs_2(as_2-2as_1-bs_1)\\as_1(bs_1-2bs_2-as_2)\end{array
}\right]\\
&&+\frac{ab(s_1-s_2)^2}{(a+b)(as_2+bs_1)^3}
\left[\begin{array}{c}1\\1\end{array}\right].
\end{eqnarray*}
Now we can proceed to find the inlet condition for the zonal model.
Assume $\vec{u}(0,t)$ is given from $c(0,t)$ by an expression of the form
\begin{equation}
\vec{u}(0,t)=\overline{u(y)\vec{\zeta}_0(y)c(0,t)}
+\overline{u(y)\vec{\zeta}_{1}(y)\D t{c}(0,t)}\,.
\label{zm:u0bc2}
\end{equation}
Compare (\ref{zm:cmbc}) with the results of the appropriate transform,
(\ref{zm:u0bc2}) followed by~(\ref{zm:u0bc}) and require that the
coefficients of $c(0,t)$ and its derivatives are equal to obtain
\begin{eqnarray*}
p_0(y)&=&\vec{P}_0^TB\vec{\zeta}_0\,,\\
p_1(y)&=&\vec{P}_0^TB\vec{\zeta}_1+\vec{P}_1^TB\vec{\zeta}_0\,,\\
p_2(y)&=&\vec{P}_1^TB\vec{\zeta}_1+\vec{P}_2^TB\vec{\zeta}_0\,,\\
p_3(y)&=&\vec{P}_2^TB\vec{\zeta}_1+\vec{P}_3^TB\vec{\zeta}_0\,.
\end{eqnarray*}
The solution to these linear equations is
\begin{eqnarray*}
\vec{\zeta}_0(y)&=&
\left[\begin{array}{c}1.9242\\-2.858\end{array}\right]p_{0}+
\left[\begin{array}{c}-20.40\\85.14\end{array}\right]p_{1}+
\left[\begin{array}{c}-1003.6\\4188\end{array}\right]p_{2}+
\left[\begin{array}{c}-14490\\60466\end{array}\right]p_{3}\,,\\
\vec{\zeta}_1(y)&=&
\left[\begin{array}{c}0.010256\\-0.2642\end{array}\right]p_{0}+
\left[\begin{array}{c}1.6112\\5.418\end{array}\right]p_{1}+
\left[\begin{array}{c}31.02\\213.4\end{array}\right]p_{2}+
\left[\begin{array}{c}229.2\\3994\end{array}\right]p_{3}\,.
\end{eqnarray*}
These functions are shown graphically in Figure~\ref{zm:bcnd}. Note that
the integral~(\ref{zm:u0bc2}) has been weighted by the advection velocity
$u(y)$. Thus they apply directly to the cross-channel distribution of the
flux of contaminant at the inlet. The dominant contribution to the fast
zone is from the centre region of the channel, the dominant positive
contribution to the slow zone is from the sides of the channel. Note the
negative contribution to the slow zone from a mid-channel injection: a
negative concentration travelling slowly in the slow zone of the model, in
effect, increases the speed and reduces dispersion of the model's
predictions---as appropriate for an injection into the fast flow and
little shear of the channel centre.
\begin{figure}
\caption{graphs of the boundary condition functions as a function of $y$
for the zonal model without diffusion: (a) ${\zeta}_0$; (b) ${\zeta}_1$.
The solid line (---------) is for the fast zone, ${\zeta}_{n1}$, and the
dotted line ($\cdots\cdots\cdots$) is for the slow zone, ${\zeta}_{n2}$.}
\label{zm:bcnd}
\end{figure}

These two prescribed inlet boundary conditions give enough boundary
conditions for the advection model to form a well posed model.

The corrections to the boundary conditions are required so that the
approximate models and original system are asymptotically equal, as described
in \cite{Rob92a}. However, if the boundary conditions of the original system
are independent of time, the boundary conditions of the zonal models are just
a weighted average of the inlet concentrations.

\subsection{Physical boundary conditions}

For the model with diffusion, given the inlet boundary conditions of the
previous section (or a modification thereof), a further two boundary
conditions are needed to form a well posed model.

Consider the zonal model
\[
\D t{\vec u}=A\vec{u}-B\D x{\vec u}+D\DD x{\vec u}\,.
\]
Converting the zonal model (\ref{zm:icz}) to a system of first order
partial differential equations with space as the time-like variable gives
\begin{eqnarray}
\D x{\vec u}&=&\vec{v}\label{zm:pde1}\,,\\
D\D x{\vec v}&=&\D t{\vec u} - A\vec{u} + B\vec{v}\,.\label{zm:pde2}
\end{eqnarray}
Substituting
\[
\left[\begin{array}{c}\vec{u}\\\vec{v}\end{array}\right]( =\vec{U} )\sim
\vec{k}e^{\lambda x}\,,
\]
into (\ref{zm:pde1}--\ref{zm:pde2}) leads to a perturbed eigenvalue problem
if the time derivative, $\D t{}$, is assumed to be a ``small''
perturbation. The eigenvalues of this system are $\lambda_1\approx 2745$,
$\lambda_2\approx 244$, $\lambda_3\approx 0$ and $\lambda_4\approx -13$.
\begin{itemize}
\item The approximate zero eigenvalue corresponds to the
slow evolution in the interior of the domain.
\item Near the inlet, there will be transients behaving like $e^{-13x}$.
These are
acceptable as they correspond to the not-so-fast relaxation from the two
inlet conditions~(\ref{zm:u0bc2}) to the slowly varying interior dynamics.
\item However, near the exit, there are two rapid
exponential transients arising from modes corresponding to the large
positive eigenvalues $\lambda_1$
and $\lambda_2$.  These modes must not be present and boundary conditions
are here found to eliminate them.
\end{itemize}

Suppose the system (\ref{zm:pde1}--\ref{zm:pde2}) has eigensolutions
\{$\lambda_k ; \vec{z}_k ; \vec{u}_k$\} where $\lambda_k$ is the
eigenvalue, $\vec{z}_k$ is the left eigenvector corresponding to
$\lambda_k$, $\vec{u}_k$ is the right eigenvector corresponding to
$\lambda_k$, and $\vec{z}_i^T \vec{u}_j=\delta_{ij}$. Now the solution at
the exit, $x=L$, is
\[
\vec{U}(L,t)=\sum_{i=1}^{4}\alpha_{i}(t)\vec{u}_i\,,
\]
where $\alpha_i(t)=\vec{z}_i^{T}\vec{U}(L,t)$.
To eliminate the modes corresponding to $\lambda_1$ and $\lambda_2$
at the exit, we thus require
\[
\left[\begin{array}{c}
\vec{z}_1^T\\\vec{z}_2^T\end{array}\right]\vec{U}=\vec{0}
\mbox{ at } x=L\,.
\]
This gives two boundary conditions in the four unknowns.

At leading order
\[
\D x{\vec{u}}\sim
\left[\begin{array}{cc}-3.2862&3.2862\\10.177&-10.177\end{array}\right]\vec{
u}\,,
\]
which asserts that any concentration difference between the fast and slow
zone must correspond to a specific spatial gradient in the zones. These
allowed spatial gradients correspond to the relatively slow dynamics,
roughly $e^{0x}$ and $e^{-13x}$, in the model which have some physical
basis.

\section{Conclusion}

We have combined some of the best features of two different approaches to
modelling shear dispersion into a single approach. The first approach is
that of using centre manifold theory to derive a generalised Taylor
description of dispersion. The advantage of this particular approach is
that it is a straightforward mechanistic process to find high order
approximations. Its disadvantage is its limited spatio-temporal resolution.
The second approach is to derive a zonal model of dispersion. Here we
developed a two-zone model, the two zones corresponding to a fast zone and
a slow zone. Previously \cite{Chik86}, the coefficients of such a model
have been obtained by heuristic arguments. Here centre manifold techniques
are used to form a description of the long-term behaviour of this model,
then by matching it is possible to find the various parameters of the
interaction, advection and diffusion in the zonal model. In essence, this
is the same principle as that employed in constructing Pad\'e
approximations of a power series.

That the principle can work is shown by the excellent agreement exhibited
in Section~3.6 between the predictions of the zonal model and the solutions
of the original system. Although we have not been able to quantify the spatial
resolution of the zonal models (as has been done for other dispersion
models\cite{MR90,WR94,MR94}), nonetheless, Figure~\ref{zm:comp} indicates that
the resolution is significantly improved.

As noted in Section~2, we reasonably
expect the exponential transients of such a zonal system to approximately
model actual physical dynamics; this is because physical transients are a
continuation of the centre manifold expansion through the complex plane by
being different branches of the one analytic function. However, in this
problem the Riemann sheets of the symmetric and asymmetric modes are
entirely disjoint---due to symmetry of the problem there is no interaction
between the two types of modes. Thus a zonal model constructed by matching
centre manifolds can never ``know'' about dynamics of the asymmetric modes,
and is thus deficient in the asymmetric dynamics. To construct a zonal
model that resolves some asymmetric dynamics, perhaps we would need to
introduce
some asymmetry in order to couple the symmetric and asymmetric modes.

By using asymptotically correct initial conditions, we compared the various
models in Section~3.6. This demonstrated the close agreement of the zonal
models with the original system at small time. By changing the view of the
evolution from temporal to spatial \cite{Rob92a}, we followed the same
matching procedure to obtain inlet conditions of the zonal model given the
inlet boundary conditions of the original system. Outlet boundary conditions
were obtained, as in \cite{Rob92a}, by requiring that there be no
unphysically rapid transients at the outlet.

This new method of matching centre manifolds allows us to systematically
derive low-dimensional models, {\em complete} with initial and boundary
conditions. Furthermore, the derivation, based on centre manifold analysis,
is significantly simpler that the comparable invariant manifold analysis
\cite{WR94}.

There are a few ways in which this model could be extended. The first is to
find the inlet condition of the zonal model with diffusion. It was not
done here as it is more complicated, and should not give any qualitatively
different results. A second is to include more general interactions such as
interzonal diffusion into the zonal model. This is possible as all we would
do is to calculate the evolution on each centre manifold to another couple
of orders. A final extension would be to introduce a third zone (as was
briefly discussed in \cite{CO85}), but this would not by itself introduce
asymmetry into the zonal model---a new idea is needed to overcome this
deficit.

\end{document}